%% file: PaperForReview.tex
\documentclass[10pt,twocolumn,letterpaper]{article}

\usepackage[pagenumbers]{cvpr} 

\usepackage{graphicx}
\usepackage{amsmath}
\usepackage{amssymb}
\usepackage{booktabs}
\usepackage{subcaption}
\usepackage{makecell}
\usepackage{multirow}
\usepackage[utf8]{inputenc}

\usepackage{siunitx}
\usepackage{bm}

\usepackage[pagebackref,breaklinks,colorlinks]{hyperref}

\usepackage[capitalize]{cleveref}
\crefname{section}{Sec.}{Secs.}
\Crefname{section}{Section}{Sections}
\Crefname{table}{Table}{Tables}
\crefname{table}{Tab.}{Tabs.}

\def\cvprPaperID{326} 
\def\confName{CVPR}
\def\confYear{2023}

\begin{document}

\title{On the Audio-visual Synchronization for Lip-to-Speech Synthesis}

\author{\textit{Zhe Niu} and \textit{Brian Mak} \\
Department of Computer Science and Engineering \\
The Hong Kong University of Science and Technology \\
{\tt\small \{{zniu,mak\}}@cse.ust.hk}
}
\maketitle

\input{sections/abs.tex}
\input{sections/intro.tex}
\input{sections/related.tex}
\input{sections/approach.tex}
\input{sections/settings.tex}
\input{sections/results.tex}
\input{sections/conclusion.tex}
\input{sections/ack.tex}

\newpage
{\small
	\bibliographystyle{ieee_fullname}
	\bibliography{egbib}
}

\end{document}

%% file: sections/abs.tex
\begin{abstract}
	Most lip-to-speech (LTS) synthesis models are trained and evaluated under the assumption that the audio-video pairs in the dataset are perfectly synchronized.
	In this work, we show that the commonly used audio-visual datasets, such as GRID, TCD-TIMIT, and Lip2Wav, can have data asynchrony issues. Training lip-to-speech with such datasets may further cause the model asynchrony issue --- that is, the generated speech and the input video are out of sync.
	To address these asynchrony issues, we propose a synchronized lip-to-speech (SLTS) model with an automatic synchronization mechanism (ASM) to correct data asynchrony and penalize model asynchrony.
	We further demonstrate the limitation of the commonly adopted evaluation metrics for LTS with asynchronous test data and introduce an audio alignment frontend before the metrics sensitive to time alignment for better evaluation.
	We compare our method with state-of-the-art approaches on conventional and time-aligned metrics to show the benefits of synchronization training.
\end{abstract}

%% file: sections/intro.tex
\section{Introduction}

Lip-to-speech (LTS) is the task of reconstructing the speech audio of a speaker based on the lip movement in a silent video.
%
%
%
With the development of deep learning, many data-driven deep network models have been proposed to solve the LTS task.

\begin{figure}[t]
	\centering
	\begin{subfigure}[b]{0.48\linewidth}
		{\includegraphics[width=\linewidth,page=1,trim={325 165 360 150},clip]{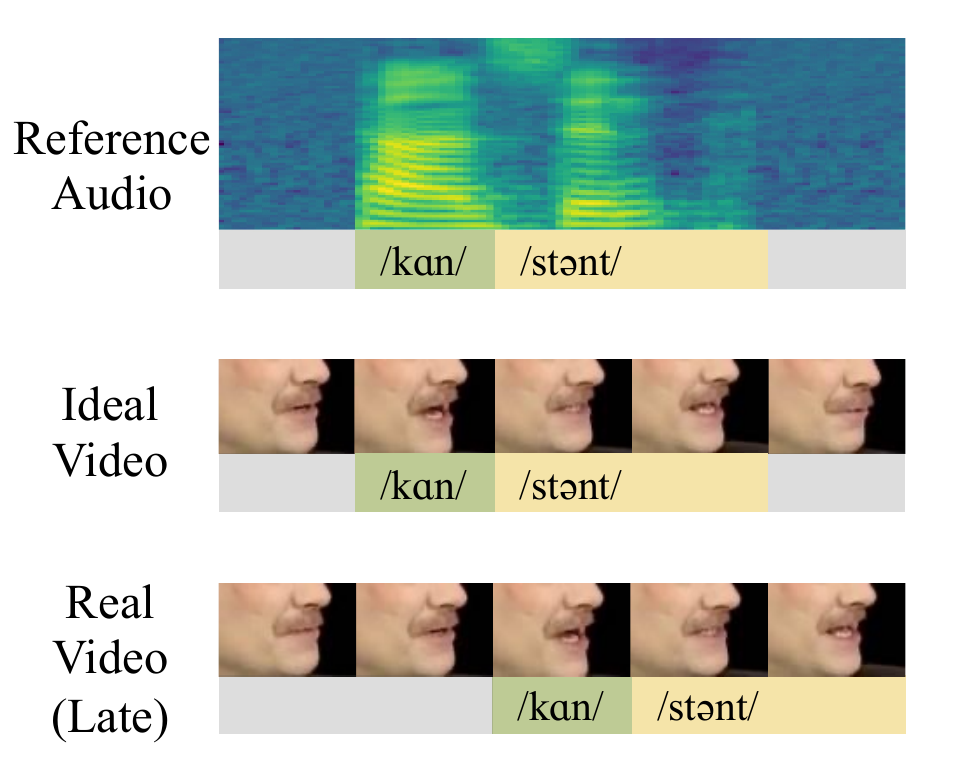}}
		\caption{Data asynchrony issue.}
		\label{fig:data-asynchrony}
	\end{subfigure}%
	\unskip\ \vrule\
	\begin{subfigure}[b]{0.48\linewidth}
		{\includegraphics[width=\linewidth,page=2,trim={325 165 360 150},clip]{figures/CVPR2023.pdf}}
		\caption{Model asynchrony issue.}
		\label{fig:model-asynchrony}
	\end{subfigure}%
	\caption{Illustration of the audio-visual asynchrony problem. The speaker is saying the word \textit{constant}.}
\end{figure}

A common assumption is made in training LTS models: the time offset between the corresponding video and audio data is a small constant, or zero. In other words, the audio and video of the same speech are fairly synchronized in time.
However, after analyzing the synchronization errors using the lip-sync model SyncNet~\cite{chung2016lipsync}, we find that there exist varying time offsets between audios and videos in the audio-visual datasets that are commonly used for training and evaluating LTS models. Some datasets, such as GRID~\cite{cooke2006audio} and TCD-TIMIT~\cite{harte2015tcd} have small offsets within $\pm1$ video frames, but others, such as Lip2Wav~\cite{prajwal2020learning}, may have larger offsets of multiple video frames.
Moreover, large time offsets can also be introduced with careless data preprocessing (\eg using FFmpeg~\cite{tomar2006converting} to segment a video file into smaller chunks\footnote{Such segmentation exists in the preprocessing pipeline of the Lip2Wav dataset: \url{https://github.com/Rudrabha/Lip2Wav/blob/master/download_speaker.sh}.}). We call this \textit{data asynchrony issue} (see \cref{fig:data-asynchrony}) since the synchronization error comes from the external dataset instead of the LTS model itself.

Although the synchronization errors are, most of the time, barely visible to the human naked eye, they can have a non-negligible impact on LTS model optimization. The training of LTS models usually utilizes time-wise learning objectives (\eg MSE between the audio mel-spectrograms) that are sensitive to time offsets. The misalignments between videos and audios in the dataset can mislead the model to produce asynchronous output, resulting in the \textit{model asynchrony issue} (see \cref{fig:model-asynchrony}).
Besides, non-constant time offsets can cause training instability, making it difficult for the model to converge on large-scale datasets.

In the evaluation stage, the audio-visual asynchrony of the test dataset can make objective evaluation difficult as well. The commonly used objective speech intelligibility measures, such as STOI~\cite{taal2010stoi} and ESTOI~\cite{jensen2016estoi}, require the reference audio and the testing audio to be perfectly time-aligned to produce scores that precisely reflect the outcome of a listening test. When both model and data asynchrony are present, misalignment between audio and video of the same speech can lead to inaccurate evaluation if they are not handled carefully.

In this work, we aim to solve the asynchrony issues in both the training and evaluation stage.
For training, we introduce the synchronized lip-to-speech (SLTS) architecture, which consists of an automatic synchronization mechanism (ASM) that ensures the model and data synchronization in the training stage.
Moreover, we propose an intrusive time-alignment frontend of the popular metrics during evaluation.
The proposed frontend decouples the synchronization errors from conventional evaluation, ensuring reliable scoring despite data asynchrony in the test set.

In the experiment section, we perform extensive experiments on popular audio-visual datasets to show the effectiveness of the proposed automatic synchronization mechanism.
The results show that the proposed synchronization method can handle both the long-term asynchrony that is visible to the human eye (\eg, more than one video frame offsets) and subtle synchronization errors (\eg, single-frame or sub-frame offsets). SLTS also outperforms existing SOTA models on various objective metrics and achieves high scores in the subjective listening test.

%% file: sections/related.tex
\section{Related Works}

\subsection{Synchronization in Lip-to-Speech Models}

Lip-to-speech models usually consist of components that provide a large temporal receptive field, such as 3D convolutional stacks~\cite{prajwal2020learning}, LSTM or GRU~\cite{akbari2018lip2audspec,prajwal2020learning,yadav2021speech,mira2022end}, location sensitive attention~\cite{prajwal2020learning,hong2021speech}, and self-attention layers~\cite{kim2021vcagan, wang2022fastlts}.
The large receptive field potentially allows the model to generate offset audio.
Kim~\etal~\cite{kim2021vcagan} point out that some existing LTS models do not explicitly process local visual features and may produce out-of-sync speech from the input video. They propose additional synchronization losses during training to handle the model asynchrony problem.
However, their work only considers the model asynchrony but not the data asynchrony.
Our work considers both types of asynchrony and proposes solutions to these issues.

\subsection{Lip-Sync Models}

The task of lip-sync aims to predict audio-visual offsets to correct lip-sync errors.
Existing works, such as~\cite{chung2016lipsync,kim2021end}, assume the audio-visual training data is synchronized and design different negative pairs to train the model with contrastive loss.
Chung~\etal~\cite{chung2016lipsync} generate negative (off-sync) audio-video pairs by randomly shifting the audio and applies the contrastive loss from Siamese networks~\cite{chopra2005learning} to train their network.
Kim~\etal~\cite{kim2021end} instead adopt a softmax-based contrastive loss and treats the audio-visual features with different time steps as negative pairs.
Our proposed data synchronization module (DSM) can also be used for lip-sync. Compared to existing lip-sync models, DSM does not assume the training data to be synchronized. It processes a set of candidate pairs and discovers the positive and negative pairs in an unsupervised manner, driven by the lip-to-speech learning objective (\eg, MSE loss between mel-spectrograms).

\subsection{End-to-End Lip-to-Speech Models}

Lip-to-speech models are often not designed to generate waveform end-to-end since more compact acoustic representations (\eg mel-spectrogram) are usually sought to reduce the task difficulty.
The compact acoustic representations are later converted to audio waveform by a vocoder, which can be either algorithm-based, such as Griffin-Lim used in~\cite{prajwal2020learning,kim2021vcagan,yadav2021speech}, or a separately trained neural vocoder as in~\cite{kim2021multi,hong2021speech,mira2022svts}.
Building end-to-end LTS models~\cite{mira2022end,wang2022fastlts} that directly generate the audio waveform has recently attracted more attention as it produces speech with better quality than the algorithm-based vocoder and does not require separate training of a neural vocoder.
In this work, we also investigate end-to-end modeling by jointly training a UnivNet vocoder~\cite{jang2021univnet} with the proposed model.

%% file: sections/approach.tex
\newcommand{\vid}{x}
\newcommand{\aud}{y}
\newcommand{\haud}{\hat{y}}

\newcommand{\dofst}{{o_d}}
\newcommand{\mofst}{{o_m}}

\newcommand{\Vid}{\bm{X}}
\newcommand{\Aud}{\bm{Y}}
\newcommand{\Feat}{\bm{F}}
\newcommand{\Mel}{\bm{M}}
\newcommand{\hMel}{\hat{\bm{M}}}
\newcommand{\hmel}{\hat{\bm{m}}}

\newcommand{\hHard}{\hat{\bm{M}}^h}
\newcommand{\hSoft}{\hat{\bm{M}}^s}
\newcommand{\hhard}{\hat{\bm{m}}^h}
\newcommand{\hsoft}{\hat{\bm{m}}^s}

\newcommand{\Eaud}{\bm{U}}
\newcommand{\Evid}{\bm{V}}

\newcommand{\eaud}{\bm{u}}
\newcommand{\evid}{\bm{v}}

\newcommand{\Corr}{\bm{c}}
\newcommand{\corr}{c}

\begin{figure*}
	\centering
	{\includegraphics[width=\linewidth,page=3,trim={15 130 5 205},clip]{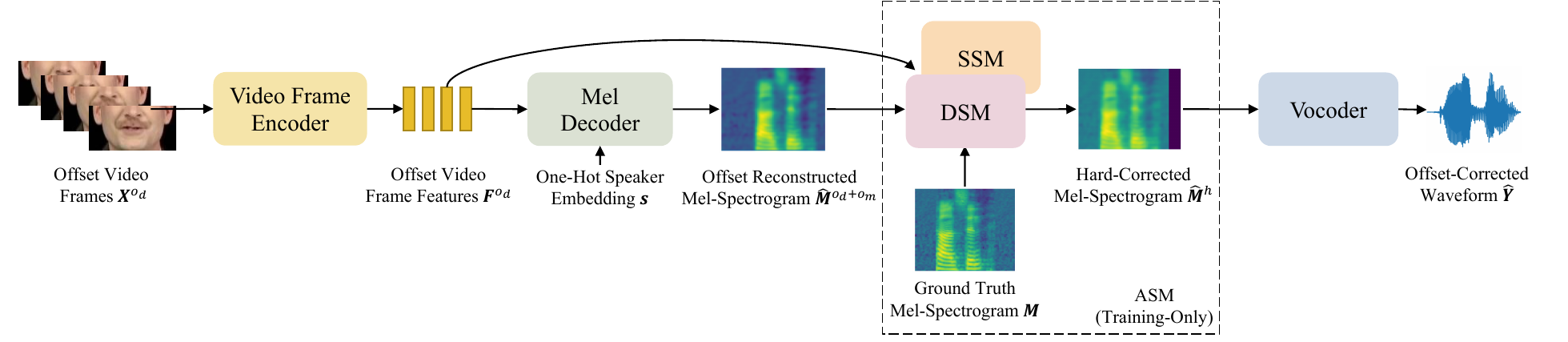}}
	\caption{The overview of the proposed SLTS architecture. Where $\dofst$ is due to data asynchrony and $\mofst$ is due to model asynchrony; both are measured in seconds. The two asynchrony issues are handled by DSM and SSM respectively.}
	\label{fig:overview}
\end{figure*}

\section{Synchronized Lip-to-Speech Synthesis}

We first formulate the data and model asynchrony issues in \cref{sec:formulation},
and then we describe the overall architecture of the proposed {synchronized lip-to-speech (SLTS)} model in \cref{sec:overview}.
We will introduce our key contribution, the {automatic synchronization mechanism (ASM)}, with a detailed description on its two components:
the {data synchronization module (DSM)} and the {self-synchronization module (SSM)} in \cref{sec:ASM}.

\subsection{Problem Formulation}
\label{sec:formulation}

For simplicity, we first consider the silent lip video ${\vid(t) \in \mathbb{R}^{H \times W \times 3}}$ and the corresponding audio ${\aud(t) \in \mathbb{R}}$ as continuous functions of time ${t \in \mathbb{R}}$. There are two kinds of asynchrony issues that LTS task faces: the data asynchrony issue and the model asynchrony issue.

Ideally, a lip video $\vid(t)$ is expected to be accompanied by an audio $\aud(t)$ with a zero time offset. However, in real-world datasets, a lip video can have a non-constant time offset of $\dofst$ seconds from the audio (when $\dofst > 0$, video lags behind the audio), resulting in an asynchronous video: ${\vid^\dofst(t) = \vid(t - \dofst)}$. We call this $\dofst$-second data asynchrony issue as shown in \cref{fig:data-asynchrony}.

On the other hand, an LTS model can inject its own time offset to the reconstructed audio due to its exploitation of temporal context when the data used to train the model is off-sync.
Given a video $\vid(t)$, the LTS model may instead reconstruct an audio ${\haud^\mofst(t) = \haud(t - \mofst)}$ with a time shift of $\mofst$ (seconds) from its ideal synchronized reconstruction $\haud(t)$. We call this the $\mofst$-second model asynchrony as shown in \cref{fig:model-asynchrony}.

Usually, if there is no data asynchrony, there may not be model asynchrony neither as the synchronized reconstructed audio should be the optimal among other asynchronous proposals. On the other hand, data asynchrony will bring forth model asynchrony, especially when the audio-visual offsets vary from samples to samples.

\subsection{Architecture Overview}
\label{sec:overview}

Before delving into the solution to the asynchrony problems, we first introduce the architecture of our LTS model, which is shown in \cref{fig:overview}.
In practice, the video and audio data are discrete signals in time.
The silent RGB video data is represented by $\Vid \in \mathbb{R}^{T_v \times H \times W \times 3}$
where $T_v, H, W$ are the number of video frames, frame height and width, respectively, and $3$ is the number of color channels.
The single-channel audio with $T_a$ samples is represented by
$\Aud \in \mathbb{R}^{T_a \times 1}$.
In our work, the video data have the frequency of $25$ or $30$ Hz depending on the dataset, and the audio frequency is fixed to $16$ kHz.

The proposed synchronized lip-to-speech (SLTS) model aims to reconstruct an offset-corrected audio $\hat{\Aud}$ from a given silent video $\Vid^\dofst$ which has an offset of $\dofst$ (seconds) during training to match the learning target (ground truth) audio $\Aud$. During inference, the model can either produce an audio $\hat{\Aud}^\dofst$ that is aligned with the offset video $\Vid^\dofst$ without using ASM, or an ASM-corrected audio $\hat{\Aud}$ aligned with the reference audio $\Aud$. The latter is mainly used for evaluation which requires aligned audios.

SLTS consists of a video frame encoder, a decoder, two synchronization modules, namely DSM and SSM, and a vocoder.
The frame encoder is based on ResNet18~\cite{he2016resnet}, which produces $D_f$-dimensional features $\Feat \in \mathbb{R}^{T_v \times D_f}$ for each individual video frame.
%
%
The decoder consists of a conformer~\cite{gulati2020conformer} and a Conv1D-based post-net.
%
The 25 Hz frame features $\Feat$ are first concatenated with the speaker embedding and then sent to the conformer to generate compact acoustic representations based on local and global contexts.
%
The compact acoustic representations are then linearly upsampled to 100 Hz and fed into the post-net to generate 100 Hz mel-spectrograms $\hMel$.

Following the decoder is the ASM which consists of two modules: DSM and SSM, the key contributions of this work. The two modules learn and correct the data and model asynchrony respectively during training. DSM takes the 25 Hz video frame features $\Feat$, the ground-truth mel-spectrogram $\Mel$ and the reconstructed mel-spectrogram $\hMel$ as inputs to estimate the time offset for correcting the asynchrony in the audio-visual data. SSM, on the other hand, generates a self-synchronization loss based on the video frame features $\Feat$ and the reconstructed mel-spectrogram $\hMel$ to penalize the model asynchrony.


Finally, an UnivNet-based~\cite{jang2021univnet} vocoder is adopted to generate the audio waveform. Since vocoders are usually trained with audio segments shorter than $1.0$ second, we perform $0.6$-second random segmentation on pairs of offset-corrected mel-spectrogram and reference audio waveform. The vocoder is trained with the whole system using multi-resolution STFT loss and a differentiable STOI loss. We also allow the option of adopting the multi-resolution spectrogram discriminator (MRSD) and the multi-period waveform discriminator (MPWD) to further improve subjective speech quality at the cost of lower objective evaluation scores.

\subsection{Automatic Synchronization Mechanism}
\label{sec:ASM}

The automatic synchronization mechanism consists of two components: DSM and SSM, both rely on a time offset predictor.
We describe the time offset predictor first and then introduce the details of DSM and SSM.

\subsubsection{Audio-visual Time Offset Predictor}

\begin{figure}
	\centering
	\includegraphics[width=\linewidth,page=4,trim={340 50 190 60},clip]{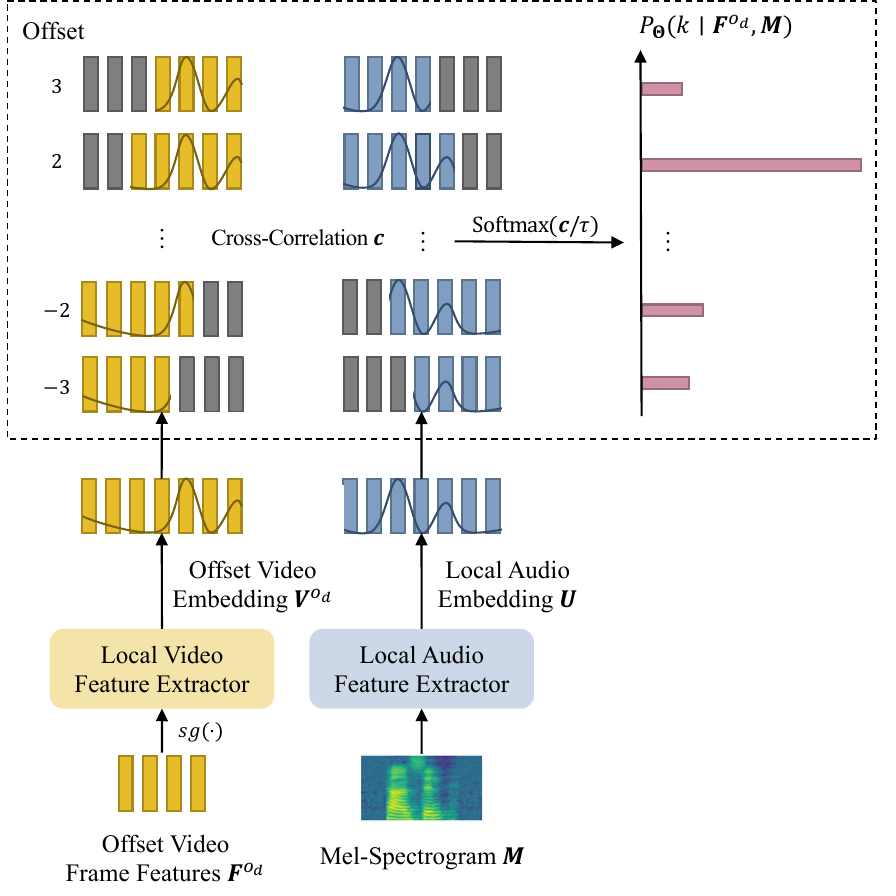}
	\caption{The audio-visual offset predictor.}
	\label{fig:op}
\end{figure}

The time offset predictor (shown in \cref{fig:op}) generates a categorical distribution for the audio-visual offsets. The range of values in the categorical distribution of time offsets can be set manually, usually from 150--300 ms.
Different from SyncNet~\cite{chung2016lipsync} which predicts synchronization error in the number of video frames (at 25 Hz in our work), the proposed offset predictor predicts offsets in the number of mel-spectrogram frames (at 100 Hz in our work) to achieve more precise synchronization.

The offset predictor contains two local feature extractors for video and audio, respectively.
Each feature extractor contains two Conv1D-BN-GELU blocks, a fully-connected layer and an L2 normalization operation to generate normalized local embeddings.
Only the first Conv1D has a kernel size of $3$, whereas the other has a kernel size of $1$.
The receptive field is intentionally restricted to preserve the time precision of the embeddings, with a slight exploitation of temporal context to improve feature discriminability.
The video feature extractor has an additional resampling operation that linearly upsamples the input video features from 25 Hz to 100 Hz before the first Conv1D, so as to match the sampling rate of mel-spectrograms.

After obtaining the sequence of local video embeddings ${\Evid^\dofst = \left(\evid^\dofst_0, \dots, \evid^\dofst_{T_m-1}\right)}$ and local audio embeddings ${\Eaud = \left(\eaud_0, \dots, \eaud_{T_m-1}\right)}$, cross-correlation ${\Corr = \left(\corr_{-K}, \dots, \corr_{K}\right)}$
between the two sequences of embeddings is computed for
samples within a synchronization radius of ${K \in \mathbb{N}^+}$ (which is a hyper-parameter):

\begin{equation}
	\corr_k = \sum_{i=\max(k, 0)}^{\min(k, 0)+T_m-1} \langle \evid^\dofst_i, \eaud_{i-k} \rangle.
\end{equation}
The cross-correlation is then normalized by the softmax function with a manually tuned temperature $\tau$ to produce the offset distribution:

\begin{equation}
	P_{\Theta}(k|\Feat^\dofst, \Mel) = \frac{\exp(\corr_{k}/\tau)}{\sum_{i=-K}^K \exp(\corr_i/\tau)},
\end{equation}
where $\Theta$ is the parameters of the offset predictor (\ie, the parameters of the local extractors).

\subsubsection{Data Synchronization Module}

\begin{figure}
	\centering
	{\includegraphics[width=0.75\linewidth,page=5,trim={350 55 323 65},clip]{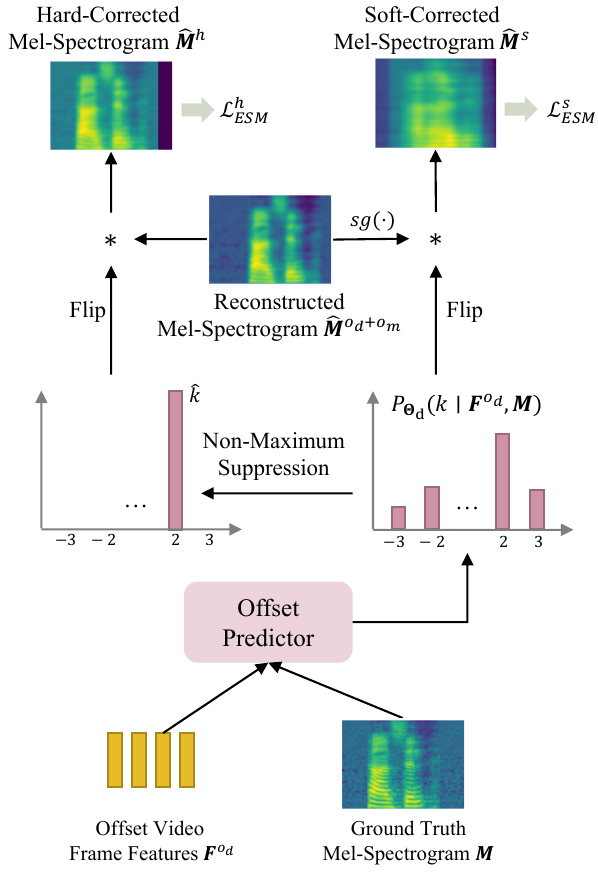}}
	\caption{The data synchronization module.}
	\label{fig:dsm}
\end{figure}

As shown in \cref{fig:dsm}, the data synchronization module (DSM) consists of an offset predictor that first generates a categorical distribution of the audio-visual offset, ${P_{\Theta_D}(k \mid \Feat^\dofst, \Mel)}$, based on the video features $\Feat^\dofst$ and the ground truth mel-spectrogram $\Mel$, with a set of DSM model parameters $\Theta_D$.

The generated offset distribution is flipped along time to obtain a correction convolution kernel, which is used to correct the offset mel-spectrogram. A soft-corrected mel-spectrogram ${\hSoft = \left(\hsoft_0, \dots, \hsoft_{T_m-1} \right)}$ is produced by convolving the reconstructed mel-spectrogram ${\hMel^{\dofst+\mofst} = \left(\hmel^{\dofst+\mofst}_0, \dots, \hmel^{\dofst+\mofst}_{T_m-1} \right)}$ with the kernel:
\begin{equation}
	\hsoft_i = \sum_{k=\max(-K, i - T + 1)}^{\min(K, i)} P_{\Theta_e}(- k \mid  \Feat^\dofst, \Mel)  sg(\hmel^{\dofst+\mofst}_{i-k}),
\end{equation}
where $sg(\cdot)$ is the gradient stopping operation.
Then, a soft-DSM loss is computed between the ground-truth mel-spectrogram and the soft-corrected mel-spectrogram:
\begin{equation}
	\mathcal{L}_{DSM}^s(\Mel, \hSoft) := \| \Mel - \hSoft \|_2^2.
\end{equation}
When generating the soft-corrected mel-spectrogram, the gradient stop operation on the reconstructed mel-spectrogram is critical.
The soft-corrected mel-spectrogram is a combination of numerous offset proposals, which may include some wrong proposals. These wrong proposals may cause erroneous gradients backpropagated to the decoder, forcing it to learn several wrong targets at once, hence causing convergence problems.

To ensure that the decoder only learns from the most probable offset proposal, alongside the soft-corrected mel-spectrogram, a hard-corrected mel-spectrogram ${\hHard = (\hhard_0, \dots, \hhard_{T-1})}$ is computed by convolving the reconstructed mel-spectrogram with another correction kernel that suppresses the less likely offsets, giving the following result:

\begin{equation}
	\hhard_i = \begin{cases} \hmel^{\dofst+\mofst}_{i-\hat{k}} ,& i \ge \hat{k}  \\ \bm{0} , & i< \hat{k} \end{cases},
\end{equation}
where ${\hat{k} = \underset{k}{\arg \max} \ P(k \mid \Feat^\dofst, \hMel)}$, and the out-of-bound frames $i < \hat{k}$ are set to zero and excluded in the loss computation. Similar to the soft-DSM loss, we adopt the MSE loss on the hard-corrected mel-spectrogram:
\begin{equation}
	\mathcal{L}_{DSM}^h(\Mel, \hHard) := \| \Mel - \hHard \|^2_2.
\end{equation}
Ideally, after the DSM is trained to convergence, a shift of $-\hat{k}$ frames on the reconstructed mel-spectrogram will correct the $\dofst$-second data asynchrony.

%

\subsubsection{Self-Synchronization Module}

Besides the $\dofst$-second data asynchrony, there can also be $\mofst$-second model asynchrony due to the large receptive field of the audio decoder. We introduce a self-synchronization module (SSM) that tries to minimize the potential model asynchrony.

Similar to the DSM, SSM also contains an independent offset predictor parameterized by $\Theta_S$ to generate an offset distribution, ${P_{\Theta_S}(k\mid \Feat^\dofst, \hMel^{\dofst+\mofst})}$. Unlike DSM, SSM focuses on reducing the offsets between the video features $\Feat^\dofst$ and the reconstructed mel-spectrogram $\hMel^{\dofst+\mofst}$ by minimizing the following SSM loss:
\begin{equation}
	\resizebox{0.99\linewidth}{!}{
		$\mathcal{L}_{SSM}(\Feat^\dofst, \hMel^{\dofst+\mofst}) := -\log P_{\Theta_S}(k=0 \mid \Feat^\dofst, \hMel^{\dofst+\mofst})$.
	}
\end{equation}

Empirically, we find that SSM improves the training stability of DSM. Without SSM,  DSM sometimes does not learn effective offsets, and the predicted offset collapses to a constant. We hypothesize that this may be attributed to the propagation of noisy gradients from the otherwise uncontrolled soft DSM loss.

%% file: sections/settings.tex
\newcommand{\fofst}{o_f}

\section{Datasets, Metrics and Training Details}

\begin{figure}[t]
	\centering
	\begin{subfigure}[b]{0.3\linewidth}
		\includegraphics[width=\linewidth,trim={22 22 22 22},clip]{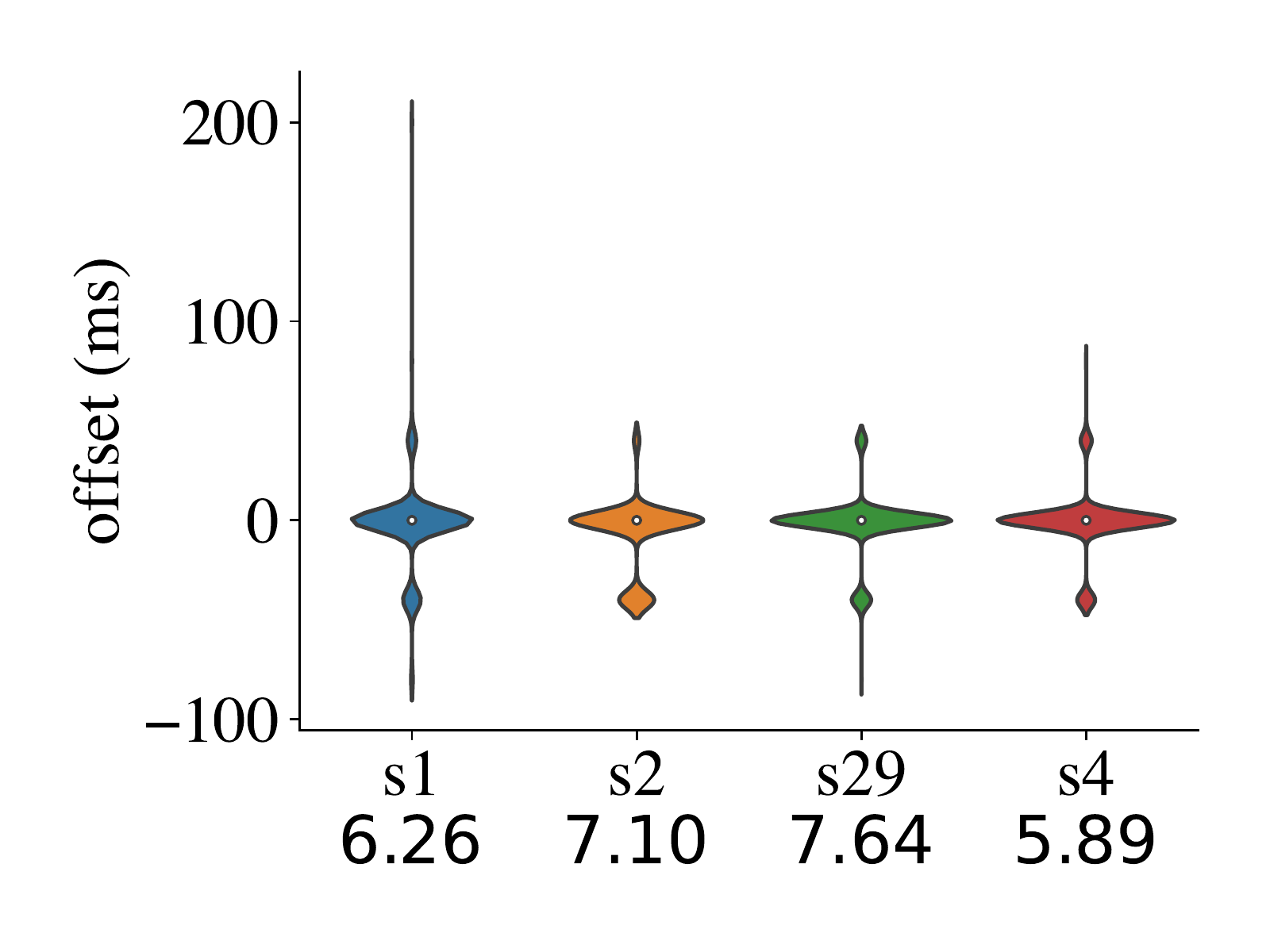}
		\caption{GRID-4S.}
	\end{subfigure}%
	~
	\begin{subfigure}[b]{0.3\linewidth}
		\includegraphics[width=\linewidth,trim={22 22 22 22},clip]{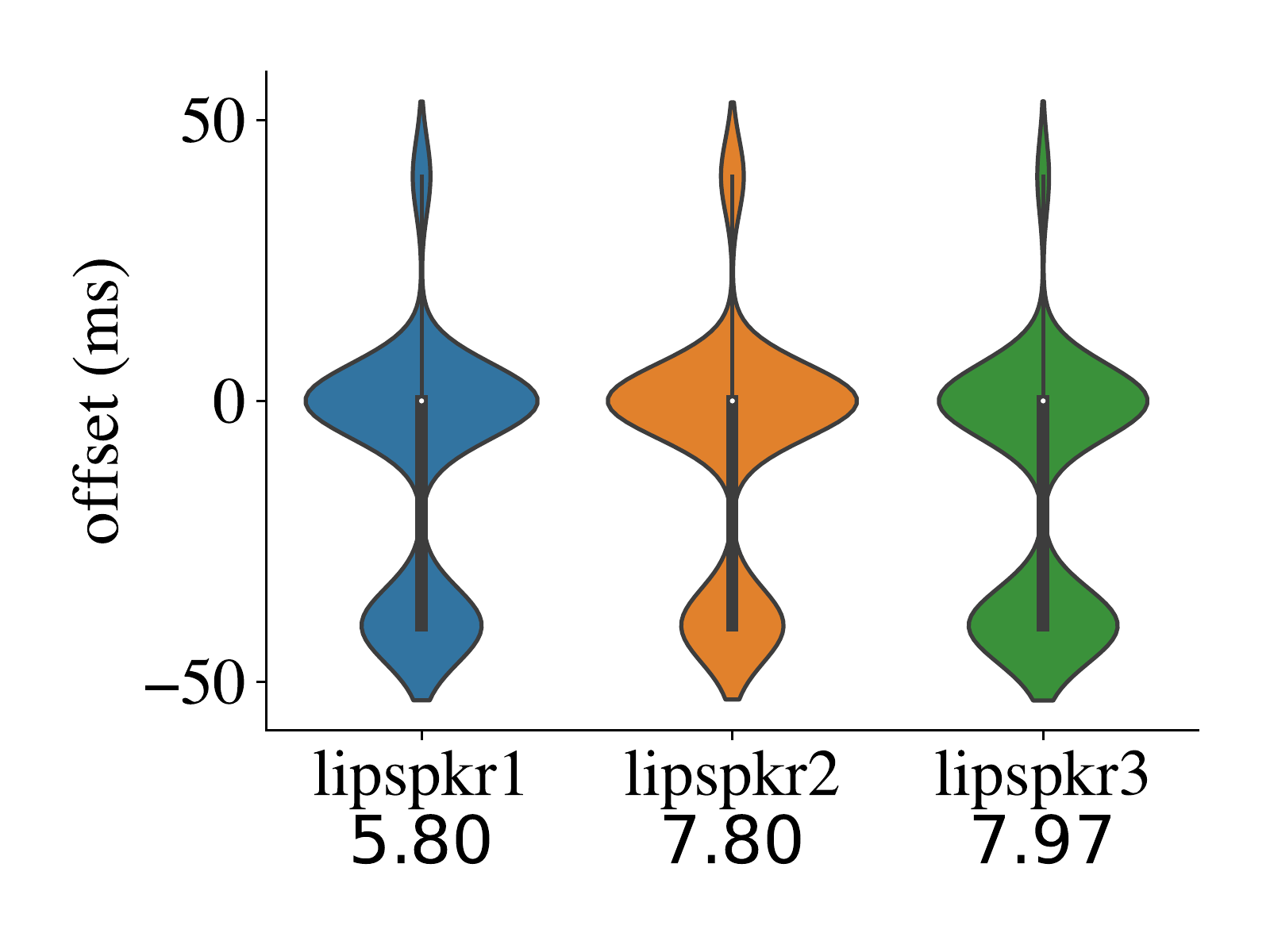}
		\caption{TCD-TIMIT-LS.}
	\end{subfigure}
	~
	\begin{subfigure}[b]{0.3\linewidth}
		\includegraphics[width=\linewidth,trim={22 22 22 22},clip]{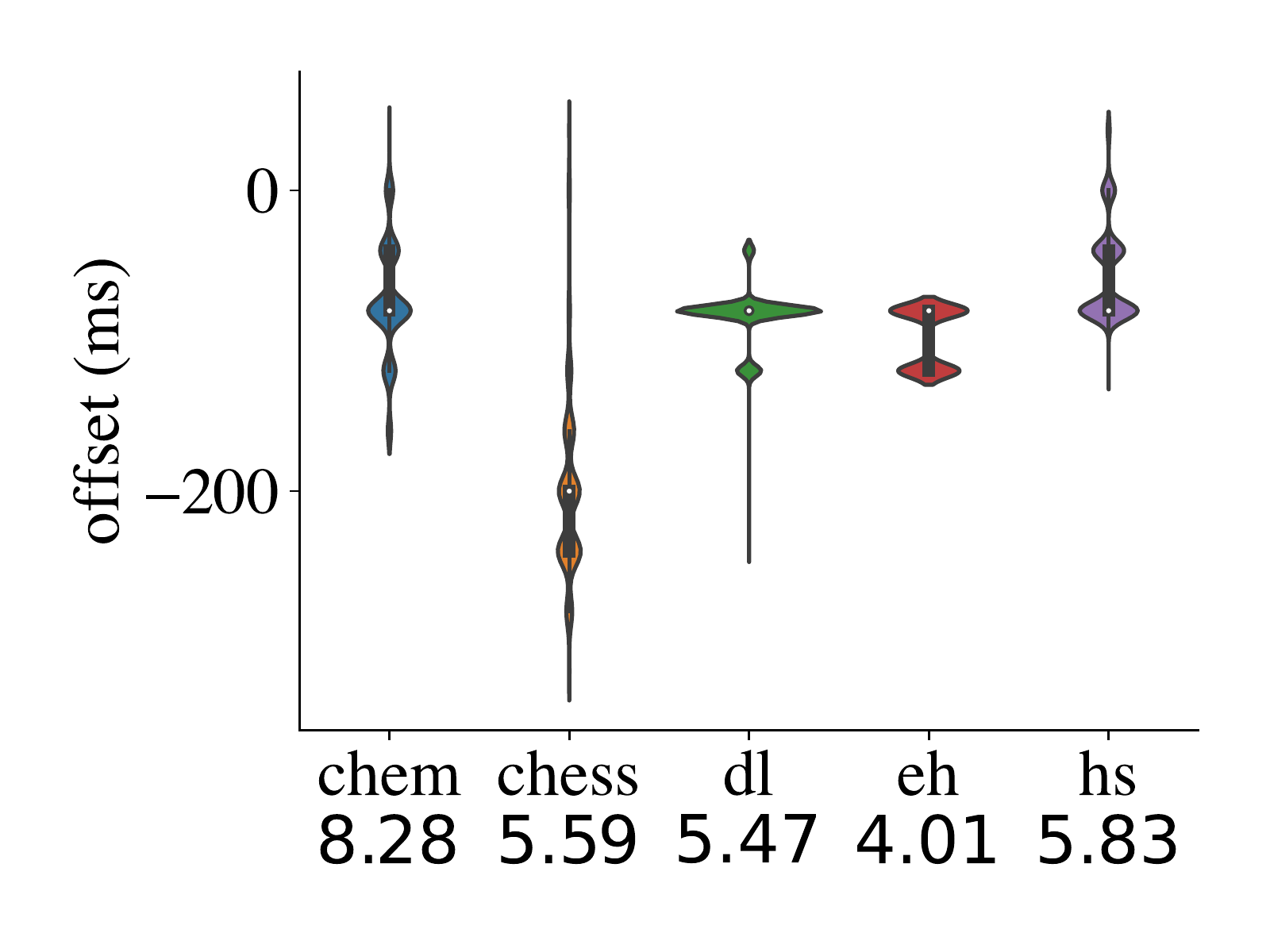}
		\caption{Lip2Wav.}
		\label{fig:async-stats-lip2wav}
	\end{subfigure}
	\caption{Offsets produced by SyncNet. The figures under speaker names are confidence scores produced by SyncNet. A higher score means SyncNet has greater confidence in its outputs. Only the offsets having confidence greater than $3.0$ are counted.}
	\label{fig:async-stats}
\end{figure}

\begin{table}[t]
	\centering
	\resizebox{\columnwidth}{!}{%
		\begin{tabular}{@{}l|c|c|c@{}}
			\toprule
			Dataset                            & Speakers & Train / Val / Test Samples & Train / Val / Test Hours \\
			\midrule
			GRID-4S \cite{cooke2006audio}      & 4        & 3,600 / 200 / 200          & 2.98 / 0.17 / 0.17       \\
			TCD-TIMIT-LS \cite{harte2015tcd}   & 3        & 1,017 / 57 / 57            & 1.64 / 0.09 / 0.09       \\
			Lip2Wav \cite{prajwal2020learning} & 5        & 15,894 / 376 / 487         & 115.16 / 2.48 / 3.37     \\
			\bottomrule
		\end{tabular}
	}
	\caption{Statistics for dataset splits used in our experiments. All speakers from the same dataset are present in all training, validation, and test splits.}
	\label{tab:dataset-stats}
\end{table}

\begin{table*}[t]
	\centering
	\resizebox{\linewidth}{!}{%
		\begin{tabular}{@{}l|l|cccc|cccc|cc|cc@{}}
			\toprule
			Dataset & Model          & STOI $\uparrow$  & ESTOI $\uparrow$ & PESQ $\uparrow$ & MCD $\downarrow$ & \textit{a-}STOI $\uparrow$ & \textit{a-}ESTOI $\uparrow$ & \textit{a-}PESQ $\uparrow$ & \textit{a-}MCD $\downarrow$ & Offset-$R^2$ (F) $\uparrow$          & Offset-$R^2$ (SN) $\uparrow$        & \textit{w-}WER (\%) $\downarrow$ & \textit{k-}WER (\%) $\downarrow$ \\
			\midrule
			\multirow{3}{*}{\makecell[l]{ GRID-4S
				}}
			        & VCA-GAN        & 0.688            & 0.500            & 1.917           & 29.720           & 0.732                      & 0.552                       & 1.910                      & 28.437                      & -0.002\textsuperscript{\textdagger}  & -0.030\textsuperscript{\textdagger} & 23.67                            & 8.50                             \\
			        & SLTS w/o ASM   & 0.698            & 0.519            & 1.906           & 27.438           & 0.753                      & 0.582                       & 1.903                      & 25.684                      & -0.001\textsuperscript{\textdagger}  & -0.030\textsuperscript{\textdagger} & 15.33                            & 4.92                             \\
			        & \bfseries SLTS & \bfseries 0.703  & \bfseries 0.525  & \bfseries 1.932 & \bfseries 27.327 & \bfseries 0.761            & \bfseries 0.592             & \bfseries 1.933            & \bfseries 25.404            & \bfseries 0.862                      & \bfseries 0.383                     & \bfseries 12.83                  & \bfseries 2.92                   \\
			\midrule
			\multirow{3}{*}{\makecell[l]{TCD-TIMIT-LS}}
			        & VCA-GAN        & 0.577            & 0.398            & 1.373           & 33.450           & 0.593                      & 0.412                       & 1.376                      & 33.175                      & -0.038\textsuperscript{\textdagger}  & -0.164\textsuperscript{\textdagger} & 79.96                            & -                                \\
			        & SLTS w/o ASM   & \bfseries 0.622  & \bfseries 0.460  & \bfseries 1.480 & \bfseries 30.334 & 0.650                      & 0.496                       & \bfseries 1.482            & 29.667                      & -0.585 \textsuperscript{\textdagger} & -0.164\textsuperscript{\textdagger} & 50.40                            & -                                \\
			        & \bfseries SLTS & 0.606            & 0.445            & \bfseries 1.480 & 30.818           & \bfseries 0.664            & \bfseries 0.511             & 1.480                      & \bfseries 29.430            & \bfseries 0.796                      & \bfseries 0.525                     & \bfseries 38.06                  & -                                \\
			\midrule
			\multirow{3}{*}{\makecell[l]{Lip2Wav
					\textit{chem}
				}}
			        & VCA-GAN        & 0.543            & 0.364            & 1.363           & 37.827           & 0.659                      & 0.477                       & 1.365                      & 34.600                      & -0.000\textsuperscript{\textdagger}  & -2.725\textsuperscript{\textdagger} & 48.20                            & -                                \\
			        & SLTS w/o ASM   & \bfseries  0.603 & \bfseries 0.445  & 1.478           & \bfseries 34.104 & 0.736                      & 0.578                       & 1.481                      & 30.291                      & -0.005 \textsuperscript{\textdagger} & -2.725\textsuperscript{\textdagger} & 33.03                            & -                                \\
			        & \bfseries SLTS & 0.215            & 0.049            & \bfseries 1.520 & 49.481           & \bfseries 0.760            & \bfseries 0.616             & \bfseries 1.515            & \bfseries 29.130            & \bfseries 0.982                      & \bfseries 0.704                     & \bfseries 24.69                  & -                                \\
			\bottomrule
		\end{tabular}
	}
	\caption{Comparison between VCA-GAN~\cite{kim2021vcagan}, SLTS without ASM during training, and SLTS.
		\textit{a-}: metrics with the time alignment frontend. \textdagger: results computed with a dummy offset predictor (\ie, always predicts 0). By default, the reference text used to compute WER is from the dataset, except for Lip2Wav, where the reference text is obtained by applying the Whisper ASR on the reference speech.}
	\label{tab:results-ours}
\end{table*}

\subsection{Datasets Overview}

\textbf{GRID-4S} is a four-speaker subset of the GRID audio-visual corpus~\cite{cooke2006audio}.
The subset consists of two male speakers (\textit{s1}, \textit{s2}) and two female speakers (\textit{s4}, \textit{s29}), and is commonly used in the literature~\cite{prajwal2020learning,kim2021vcagan} to evaluate lip-to-speech models.
The corpus is recorded in the laboratory condition. It has a small vocabulary and an artificial grammar.

\textbf{TCD-TIMIT-LS}~\cite{harte2015tcd} is another audio-visual corpus produced in the laboratory condition using real English sentences with a larger vocabulary. The original TCD-TIMIT dataset is produced by three professionally-trained lip speakers and 59 normal-speaking volunteers. Following the literature~\cite{prajwal2020learning,kim2021vcagan}, we adopt only the data from the three professionally-trained lip speakers.

\textbf{Lip2Wav}~\cite{prajwal2020learning} is a large-scale audio-visual dataset collected from YouTube lecture videos. The dataset includes five different speakers, all of whom are used in our experiments.

\subsubsection{Data Preparation}

For GRID-4S and TCD-TIMIT-LS datasets, we follow the convention~\cite{prajwal2020learning,michelsanti2020vocoderbased,vougioukas2019video} and randomly select $90\%$ of the data samples from each speaker for training, $5\%$ for validation, and $5\%$ for testing.
For Lip2Wav, we adopt the official data split\footnote{Official Lip2Wav splits: \url{https://github.com/Rudrabha/Lip2Wav/tree/master/Dataset}.}.
We adopt S$^3$FD~\cite{zhang2017s3fd} face detector to obtain the facial region of the videos for all three datasets.
Before face detection, the long videos in the Lip2Wav datasets are segmented into chunks with a maximum duration of 30 seconds, following the official Lip2Wav preprocessing pipeline.
The statistics of the preprocessed datasets are shown in \cref{tab:dataset-stats}.

\subsubsection{Asynchrony Analysis}

To study the asynchrony in the datasets, we use a pretrained SyncNet\footnote{Implementation and model checkpoint of SyncNet obtained here: \url{https://github.com/joonson/syncnet_python}.}~\cite{chung2016lipsync} to estimate the degree of asynchrony on the three audio-visual datasets.
SyncNet takes a 25 FPS video and 16 kHz audio as inputs and produces the audio-visual offsets with a resolution of 40 ms. The statistics of the SyncNet results are shown in \cref{fig:async-stats}.
Except for a few outliers, the audio-visual offsets of the GRID-4S and TCD-TIMIT-LS data samples center around 0~ms, with some slightly off-sync by one video frame (\ie, $\pm40$~ms).
In the Lip2Wav dataset, offsets of the \textit{chess} speaker center around -200$\sim$-250~ms, while offsets from other speakers center around -80~ms.
The audio lag of Lip2Wav data is mainly caused by video segmentation during data preprocessing, except for those from the \textit{chess} speaker, whose original videos are generally ahead of time.

\subsection{Evaluation Metrics Overview}


\textbf{PESQ}~\cite{rix2001perceptual}: evaluates the perceptual quality of a generated speech compared to a clean reference speech.
%
%
We follow \cite{prajwal2020learning,kim2021vcagan} to report the narrow-band MOS-LQO score of PESQ.

\textbf{STOI~\cite{taal2010stoi} \& ESTOI~\cite{jensen2016estoi}}: predicts the results of intelligibility listening tests based on the correlation of the short-time temporal envelopes between the generated and clean speech. Both metrics assume that the audios are time-aligned.

\textbf{MCD}: is another alignment-sensitive metric that measures the differences between two sequences of mel cepstra extracted from the generated audio and reference audio.

\textbf{WER}: counts the word errors in the transcriptions of the generated audios.
We use the medium version of Whisper~\cite{radford2022robust} to obtain the transcriptions, and the WER computed from them is denoted as \textit{w-}WER.
GRID-4S utterances are generated by an artificial grammar with a constrained vocabulary. Whisper, however, is a general large-vocabulary speech recognizer. It produces a lot of homophones (\eg, `red' $\rightarrow$ `read', `blue' $\rightarrow$ `blew') on GRID-4S which are counted as errors, resulting in inaccurate evaluation.
Thus, we train an ad-hoc Kaldi ASR model~\cite{PoveyASRU2011kaldi} with the GRID-4S training set to recognize the generated GRID-4S audios, and denote the resulting WER as \textit{k-}WER.

\textbf{Offset-$\bm{R^2}$}: is the coefficient of determination between the offsets produced by DSM and another approach, such as the metrics frontend (see \cref{sec:frontend}).  It is denoted as \mbox{Offset-$R^2$ (F)} and \mbox{Offset-$R^2$ (SN)}  if the other approach is the metrics frontend and SyncNet, respectively.

\subsection{Time Alignment Metric Frontend}
\label{sec:frontend}

Alignment-sensitive metrics, such as STOI, ESTOI, and MCD, can produce inaccurate scores when the two input audios are not time-aligned (details discussed in \cref{sec:metric-limitation}).
We propose a time alignment frontend to address the issue by first computing mel-spectrograms from both the generated and reference audios with a window size of {640} and a hop length of {160}, and then normalizing them along the channel dimension. Sixty-one alignment proposals are then created by shifting the generated audio from $-300$ ms to $300$ ms with a step size of $10$ ms. The shift that produces the minimum mean squared error between the two normalized mel-spectrograms is selected to correct the generated audio before scoring. The negative of the optimal shift is called the front-end offset, denoted as $\fofst$.

\subsection{Implementation and Training Details}

We limit the video clip length to a maximum of {3} seconds via random chunking and adopt a batch size of 32 to train our SLTS models.
Adam optimizer~\cite{kingma2014adam} with a linear warm-up and cosine annealing learning rate is adopted, where the number of warm-up steps is 1k and the maximum learning rate of \num{5e-4}.
We choose conformer~(S) for GRID and TCD-TIMIT models and conformer~(M) for Lip2Wav models.
All SLTS models are trained for a maximum of {50}k iterations (each taking around {1} day on an RTX 2080-Ti). To fit the model into the VRAM, we adopt the gradient checkpointing~\cite{chen2021wavegrad} on the frame encoder to reduce VRAM consumption.
For comparison, we also train the SOTA VCA-GAN model~\cite{kim2021vcagan} for a maximum number of {70}k iterations with the Adam optimizer and a fixed learning rate of \num{1e-4}.
A smaller batch size of {24} is adopted due to the larger memory consumption of the model.

%% file: sections/results.tex
\section{Experimental Results and Discussion}

Unless otherwise stated, the reported results are obtained from models with the best time-aligned STOI scores on the validation set throughout training. The time-aligned STOI is computed after every 1k iterations for GRID-4S and TCD-TIMIT-LS, and 5k iterations for Lip2Wav.
The results are computed on the original test set without applying an additional lip-sync method by default.

\subsection{Effectiveness of Synchronization Training}

Regardless of the severity of the asynchrony problems in the datasets, SLTS models score higher than its non-synchronized competing models (\ie, VCA-GAN and SLTS without ASM) according to the time-aligned metrics (see~\cref{tab:results-ours}). The results show that the synchronization training benefits the speech intelligibility, perceptual quality, and mel cepstra similarity of the reconstructed audios when appropriately evaluated. Moreover, the content correctness of the reconstructed audio is also improved with synchronization training, measured by WER.
Compared to the GRID-4S and TCD-TIMIT-LS datasets, the Lip2Wav~\textit{chem} dataset, which has a more severe data asynchrony issue, achieves a more significant performance gain, especially on intelligibility and content correctness. This suggests that the severe asynchrony in the dataset does not only produce off-sync generated audios but also the quality of the generated speech.

\subsection{Limitation of Non-aligned Metrics}
\label{sec:metric-limitation}

\begin{table}[t]
	\centering
	\resizebox{0.45\linewidth}{!}{%
		\begin{tabular}{@{}l|ccc@{}}
			\toprule
			Offset & STOI $\uparrow$ & ESTOI $\uparrow$ & MCD   $\downarrow$ \\
			\midrule
			0  ms  & 1.000           & 1.000            & 0.000              \\
			4  ms  & 0.916           & 0.869            & 11.651             \\
			8  ms  & 0.770           & 0.708            & 15.605             \\
			12 ms  & 0.660           & 0.594            & 18.712             \\
			\bottomrule
		\end{tabular}
	}
	\caption{The impact of offsets on the alignment-sensitive metrics. The results are computed on two copies of the same audio (bbaf2n from GRID-4S) with one shifted to simulate asynchrony.}
	\label{tab:sensitive}
\end{table}

Though models trained with the ASM achieve better intelligibility, perceptual quality, and content correctness as measured  by the alignment insensitive metrics (\eg, PESQ and WER) and the metrics with alignment frontend, they may not consistently score better on vanilla STOI, ESTOI, and MCD.
For these metrics, we notice that a slight offset between the testing and reference audios can have a large negative impact (see~\cref{tab:sensitive}).
The problem is that our SLTS model is trained to correct data asynchrony in the training data,
when it is used for testing, if the reference test audio and test video are off-sync, the audio reconstructed by our SLTS model from the test video (perhaps with a perfect zero offset) will also be off-sync with the reference test audio.
As a result, the SLTS models score lower on STOI, ESTOI and MCD.
This shows the limitation of the alignment-sensitive metrics. Without proper alignment, lower scores can be produced even for a better performing model.

\subsection{Accuracy of the Data Synchronization Module}

Since there are no available ground truths for the audio-visual offsets in the test set, we evaluate the accuracy of DSM by comparing the offsets produced by different approaches.
$R^2$ scores between the offsets predicted by DSM and the metrics frontend or SyncNet are also shown in ~\cref{tab:results-ours}.
The high $R^2$ scores between the offsets predicted by DSM (\ie, $\hat{o}_d$) and the metrics frontend (\ie, $\fofst = \dofst + \mofst, \mofst \approx 0$) show that DSM can predict the data asynchrony $\dofst$ accurately.
%
On the other hand, the offsets produced by DSM can explain more variance of the SyncNet offsets than a dummy offset predictor that assumes all offsets to be 0.
On the datasets with a more severe data asynchrony issue (\eg, Lip2Wav \textit{chem}), the $R^2$ score becomes more prominent due to the high total variance of the audio-visual offsets.
%

%
%
%
%

\subsection{Impact of Discriminators on Vocoder}

\begin{table}[t]
	\centering
	\resizebox{\linewidth}{!}{
		\begin{tabular}{@{}l|cccc|c|cc@{}}
			\toprule
			Method               & \textit{a-}STOI $\uparrow$ & \textit{a-}ESTOI $\uparrow$ & \textit{a-}PESQ $\uparrow$ & \textit{a-}MCD $\downarrow$ & \textit{w-}WER (\%) & MOS (I)  $\downarrow$        $\uparrow$ & MOS (N)  $\uparrow$         \\
			\midrule
			%
			VCA-GAN              & 0.659                      & 0.477                       & 1.365                      & 34.600                      & 48.20               & $3.250 \pm 0.225$                       & $2.042 \pm 0.179$           \\
			SLTS                 & \bfseries 0.760            & \bfseries 0.616             & \bfseries 1.515            & \bfseries 29.130            & \bfseries 24.69     & $3.633 \pm 0.228$                       & $1.858 \pm 0.171$           \\
			SLTS \textit{w/~dis} & 0.738                      & 0.583                       & 1.405                      & 31.856                      & 26.55               & \bfseries 4.483 $\pm$ 0.139             & \bfseries 4.267 $\pm$ 0.153 \\
			\midrule
			Real Voice           & 1.000                      & 1.000                       & 4.549                      & 0.000                       & 0.00                & $4.808 \pm 0.100$                       & $4.975 \pm 0.028$           \\
			\bottomrule
		\end{tabular}
	}
	\caption{Results on Lip2Wav \textit{chem}. \textit{w/~dis}: trained with discriminators to generate audio waveforms. MOS scores are listed with their 95\% confidence interval computed from their t-distribution.}
	\label{tab:discriminator}
\end{table}

To demonstrate the superiority of the model trained with discriminators (\ie, MRSD and MPWD) on audio generation, we conduct mean opinion score (MOS) tests by asking 12 volunteers to score 10 samples randomly selected from the Lip2Wav \textit{chem} test set. Both intelligibility and naturalness are assessed. Each volunteer rates four versions (\ie, VCA-GAN, SLTS, SLTS w/ dis, and real voice) of the 10 samples. Results in~\cref{tab:discriminator} show significant MOS gains on both intelligibility and naturalness after adopting the discriminators.
However, we notice that the objective scores are lower on both training and test samples after adopting the discriminators.
For instance, after including the discriminators, the \textit{a-}STOI score drops from 0.760 to 0.738 on the test set of Lip2Wav \textit{chem}, and from 0.855 to 0.825 on a training subset of 200 samples.
%
%
We hypothesize that the lower objective scores are caused by the non-intrusive nature of GAN training.
The discriminators encourage the generated audios to match a distribution of real audios rather than the corresponding target audios, rendering it harder to meet the intrusive learning objectives, such as MSE, multi-resolution STFT, and STOI, and resulting in lower scores on intrusive metrics.

\subsection{Qualitative Study}

\begin{figure}
	\centering
	\begin{subfigure}[b]{0.5\linewidth}
		{\includegraphics[width=\linewidth,trim={20 10 20 10},clip]{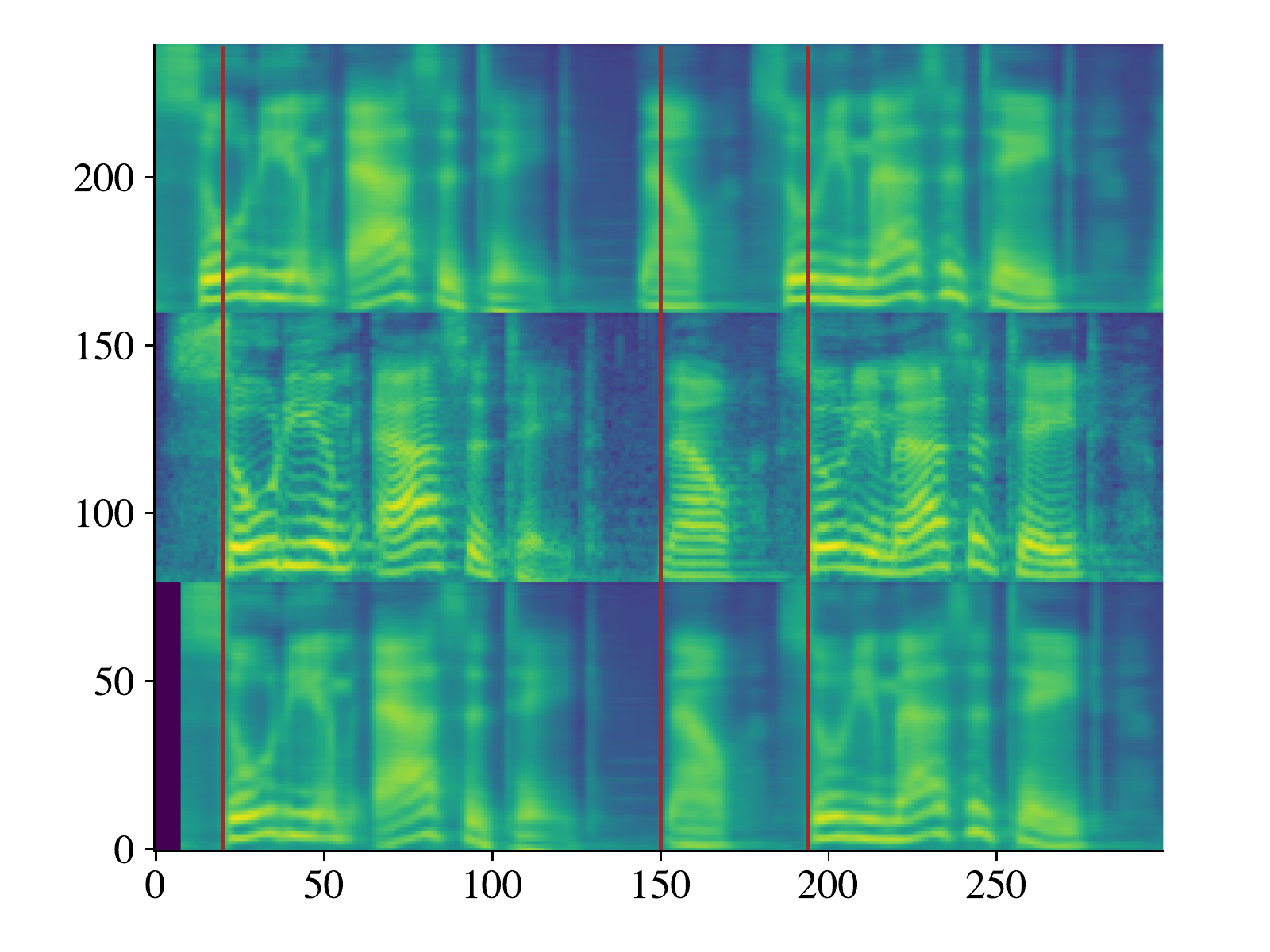}}
	\end{subfigure}
	\begin{minipage}[b]{0.25\linewidth}
		\centering
		\begin{subfigure}[b]{0.95\linewidth}
			{\includegraphics[width=\linewidth,trim={60 30 20 0},clip]{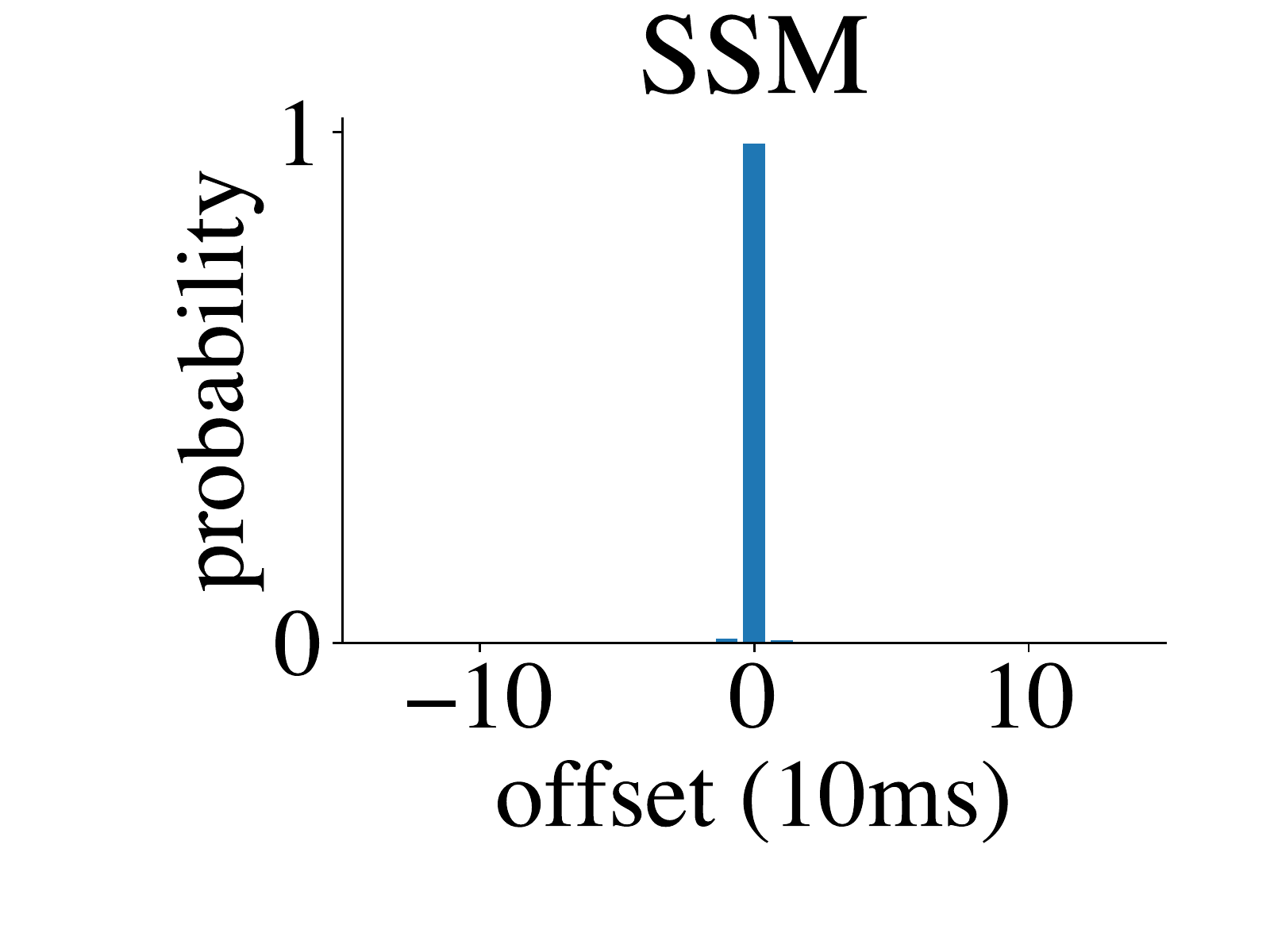}}
		\end{subfigure}
		\vfill
		\begin{subfigure}[b]{\linewidth}
			{\includegraphics[width=\linewidth,trim={11 40 30 0},clip]{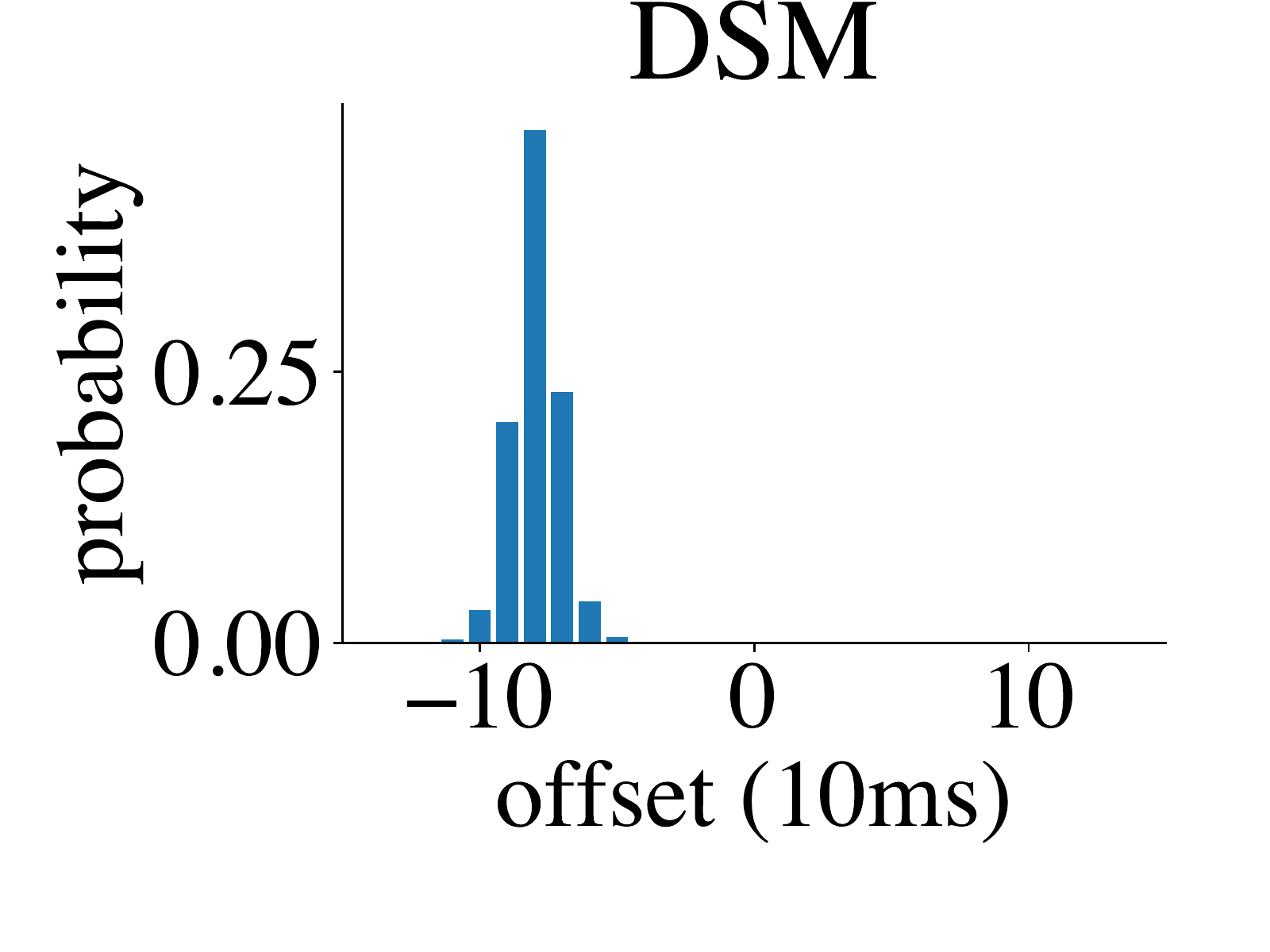}}
		\end{subfigure}
	\end{minipage}
	\caption{An example from Lip2Wav \textit{chem} showing how ASM works. The left side shows the reconstructed, ground-truth and hard-corrected mel-spectrograms from top to bottom.}
	\label{fig:demo}
\end{figure}

To demonstrate how ASM works, we show a real training example in~\cref{fig:demo}. In this example, the reconstructed audio is earlier than the reference mel-spectrogram by 80~ms. DSM assigns most of the probability mass to the offsets around -80 ms. After the reconstructed mel-spectrogram is convolved with the hard-correction kernel, the resulting mel-spectrogram precisely aligns with the ground-truth mel-spectrogram, allowing more accurate loss computation between the reconstructed and reference mel-spectrograms.

\subsection{Comparison with Other SOTA Results}

\Cref{tab:grid4s-sota,tab:tcdtimit-sota,tab:lip2wav-sota} compare our results with SOTA results reported in existing work.
Since SLTS models produce audio synchronized with the input video, they can have low scores on the vanilla metrics when data asynchrony in the test set is severe (\eg, the \textit{chem}, as shown in \cref{tab:results-ours}). For reasonable comparisons, we report the results on the test set that is lip-synced by the DSM of the corresponding SLTS model.
SLTS achieves similar or superior results compared to other SOTA works.
On GRID-4S, SLTS has the best STOI and MCD,
and outperforms other methods on all metrics except for the PESQ of work~\cite{saleem2022e2e} on TCD-TIMIT-LS.
For the majority of the speakers in Lip2Wav (\ie, \textit{chem}, \textit{chess} and \textit{hs}), SLTS achieves much better intelligibility, and comparable (or superior) perceptual quality.
On \textit{dl} and \textit{eh}, SLTS performs similarly or slightly worse than the SOTA work~\cite{hong2021speech}.
We notice that videos from \textit{dl} and \textit{eh} have relatively smaller mouth regions, making recognition of visemes difficult.
This agrees with the SyncNet results (\cref{fig:async-stats-lip2wav}) which also has lower confidence in its performance on \textit{dl} and \textit{eh}.
%

\begin{table}[t]
	\centering
	\resizebox{0.8\columnwidth}{!}{%
		\begin{tabular}{@{}l|cccc@{}}
			\toprule
			Method                                        & STOI $\uparrow$ & ESTOI $\uparrow$ & PESQ $\uparrow$ & MCD  $\downarrow$ \\
			\midrule
			E2E-V2AResNet~\cite{saleem2022e2e}            & 0.627           & -                & 2.030           & 27.790            \\
			Yadav \etal~\cite{yadav2021speech}            & 0.724           & 0.540            & 1.932           & -                 \\
			VCA-GAN~\cite{kim2021vcagan}                  & 0.724           & \bfseries 0.609  & \bfseries 2.008 & -                 \\
			Lip2Wav~\cite{prajwal2020learning}            & 0.731           & 0.535            & 1.722           & -                 \\
			Kim \etal~\cite{kim2021multi, hong2021speech} & 0.738           & 0.579            & 1.984           & -                 \\
			\midrule
			\bfseries SLTS                                & \bfseries 0.757 & 0.588            & 1.931           & \bfseries 25.491  \\
			\bottomrule
		\end{tabular}
	}
	\caption{Comparison between SOTA results on GRID-4S dataset.}
	\label{tab:grid4s-sota}
\end{table}

\begin{table}[t]
	\centering
	\resizebox{0.8\columnwidth}{!}{%
		\begin{tabular}{@{}l|cccc@{}}
			\toprule
			Method                                  & STOI           $\uparrow$ & ESTOI $\uparrow$ & PESQ         $\uparrow$ & MCD $\downarrow$ \\
			\midrule
			E2E-V2AResNet~\cite{saleem2022e2e}      & 0.472                     & -                & \bfseries 1.540         & 36.190           \\
			Ephrat \etal  \cite{ephrat2017improved} & 0.487                     & 0.310            & 1.231                   & -                \\
			GAN-based \cite{vougioukas2019video}    & 0.511                     & 0.321            & 1.218                   & -                \\
			Lip2Wav~\cite{prajwal2020learning}      & 0.558                     & 0.365            & 1.350                   & -                \\
			VCA-GAN~\cite{kim2021vcagan}            & 0.584                     & 0.401            & 1.425                   & -                \\
			\midrule
			\bfseries SLTS                          & \bfseries 0.661           & \bfseries 0.507  & 1.474                   & \bfseries 29.689 \\
			\bottomrule
		\end{tabular}
	}
	\caption{Comparison with SOTA results on TCD-TIMIT-LS.}
	\label{tab:tcdtimit-sota}
\end{table}

\begin{table}[t]
	\centering
	\resizebox{0.79\columnwidth}{!}{%
		\begin{tabular}{@{}l|l|ccc@{}}
			\toprule
			Speaker & Method                             & STOI $\uparrow$ & ESTOI $\uparrow$  & PESQ $\uparrow$ \\
			\midrule
			\multirow{3}{*}{\makecell[l]{
					\textit{chem} }}

			        & Lip2Wav \cite{prajwal2020learning} & 0.416           & 0.284             & 1.300           \\
			        & Hong \etal \cite{hong2021speech}   & 0.566           & 0.429             & \bfseries 1.529 \\
			        & \bfseries SLTS                     & \bfseries 0.757 & \bfseries   0.612 & 1.514           \\
			\midrule
			\multirow{3}{*}{\makecell[l]{\textit{chess}
				}}
			        & Lip2Wav \cite{prajwal2020learning} & 0.418           & 0.290             & 1.400           \\
			        & Hong \etal \cite{hong2021speech}   & 0.506           & 0.334             & 1.503           \\
			        & \bfseries SLTS                     & \bfseries 0.680 & \bfseries 0.451   & \bfseries 1.604 \\ 
			\midrule
			\multirow{3}{*}{\makecell[l]{\textit{dl}
				}}
			        & Lip2Wav \cite{prajwal2020learning} & 0.282           & 0.183             & \bfseries 1.671 \\
			        & Hong \etal \cite{hong2021speech}   & \bfseries 0.576 & \bfseries 0.402   & 1.612           \\
			        & \bfseries SLTS                     & 0.565           & 0.320             & 1.513           \\ 
			\midrule
			\multirow{3}{*}{\makecell[l]{\textit{hs}
				}}
			        & Lip2Wav \cite{prajwal2020learning} & 0.446           & 0.311             & 1.290           \\
			        & Hong \etal \cite{hong2021speech}   & 0.504           & 0.337             & 1.366           \\
			        & \bfseries SLTS                     & \bfseries 0.590 & \bfseries 0.394   & \bfseries 1.402 \\ 
			\midrule
			\multirow{3}{*}{\makecell[l]{\textit{eh}}}
			        & Lip2Wav \cite{prajwal2020learning} & 0.369           & 0.220             & 1.367           \\
			        & Hong \etal \cite{hong2021speech}   & 0.463           & \bfseries 0.304   & 1.362           \\
			        & \bfseries SLTS                     & \bfseries 0.482 & 0.268             & \bfseries 1.428 \\ 
			\bottomrule
		\end{tabular}
	}
	\caption{Comparison between the SOTA work and the proposed model on the Lip2Wav dataset. Unlike other datasets, we follow convention to train speaker-specific models for each speaker.}
	\label{tab:lip2wav-sota}
\end{table}

%% file: sections/conclusion.tex
\section{Conclusion}

In this work, we have identified two types of asynchronies that occur during lip-to-speech synthesis training: data asynchrony and model asynchrony. To address these asynchronies, we propose a synchronized lip-to-speech model (SLTS).
During training, the SLTS actively learns audio-visual time offsets to correct data asynchrony via DSM. The model synchronization is also ensured using SSM.
In addition, we have introduced a time alignment frontend that separates the evaluation of synchronization and audio quality from conventional time-alignment sensitive metrics, such as STOI, ESTOI, and MCD.
We have conducted extensive experiments using these new metrics to demonstrate the advantages of the proposed model. Our method achieves comparable or superior results across multiple tasks compared to existing state-of-the-art works.

%% file: sections/ack.tex
%